\documentclass[%
 reprint,
superscriptaddress,
 amsmath,amssymb,
 aps,
]{revtex4-2}

\usepackage{graphicx}
\usepackage{dcolumn}
\usepackage{bm}
\usepackage{hyperref}
\usepackage{upgreek}
\usepackage{amsmath}
\usepackage{multirow}

\begin{document}

\newcommand{\mat}[1]{\boldsymbol{#1}}

\preprint{APS/123-QED}

\title{Practical considerations for assignment of photon numbers with SNSPDs}

\author{Timon Schapeler}
\email{timon.schapeler@upb.de}
\affiliation{Department of Physics, Paderborn University, Warburger Str. 100, 33098 Paderborn, Germany}
\affiliation{Institute for Photonic Quantum Systems (PhoQS), Paderborn University, Warburger Str. 100, 33098 Paderborn, Germany}

\author{Isabell Mischke}
\affiliation{Department of Physics, Paderborn University, Warburger Str. 100, 33098 Paderborn, Germany}

\author{Fabian Schlue}
\affiliation{Integrated Quantum Optics Group, Institute for Photonic Quantum Systems (PhoQS), Paderborn University, Warburger Str. 100, 33098 Paderborn, Germany}

\author{Michael Stefszky}
\affiliation{Integrated Quantum Optics Group, Institute for Photonic Quantum Systems (PhoQS), Paderborn University, Warburger Str. 100, 33098 Paderborn, Germany}

\author{Benjamin Brecht}
\affiliation{Integrated Quantum Optics Group, Institute for Photonic Quantum Systems (PhoQS), Paderborn University, Warburger Str. 100, 33098 Paderborn, Germany}

\author{Christine Silberhorn}
\affiliation{Integrated Quantum Optics Group, Institute for Photonic Quantum Systems (PhoQS), Paderborn University, Warburger Str. 100, 33098 Paderborn, Germany}

\author{Tim J. Bartley}
\affiliation{Department of Physics, Paderborn University, Warburger Str. 100, 33098 Paderborn, Germany}
\affiliation{Institute for Photonic Quantum Systems (PhoQS), Paderborn University, Warburger Str. 100, 33098 Paderborn, Germany}

\date{\today}

\begin{abstract}
Superconducting nanowire single-photon detectors (SNSPDs) can enable photon-number resolution (PNR) based on accurate measurements of the detector's response time to few-photon optical pulses. In this work we investigate the impact of the optical pulse shape and duration on the accuracy of this method. We find that Gaussian temporal pulse shapes yield cleaner arrival-time histograms, and thus more accurate PNR, compared to bandpass-filtered pulses of equal bandwidth. For low system jitter and an optical pulse duration comparable to the other jitter contributions, photon numbers can be discriminated in our system with a commercial SNSPD. At $60~\mathrm{ps}$ optical pulse duration, photon-number discrimination is significantly reduced. Furthermore, we highlight the importance of using the correct arrival-time histogram model when analyzing photon-number assignment. Using exponentially-modified Gaussian (EMG) distributions, instead of the commonly used Gaussian distributions, we can more accurately determine photon-number misidentification probabilities. Finally, we reconstruct the positive operator-valued measures (POVMs) of the detector, revealing sharp features which indicate the intrinsic PNR capabilities.
\end{abstract}

\maketitle


\section{Introduction}\label{sec:intro}
Photon-number-resolving detectors are indispensable tools for advancing photonic quantum information science and technology, being at the heart of many experiments in fields including quantum communication~\cite{gisin2002quantum,gisin2007quantum}, quantum computing~\cite{kok2007linear}, quantum imaging and sensing~\cite{meda2017photon}, and quantum metrology~\cite{slussarenko2017unconditional}.
These detectors are crucial components to characterize quantum states~\cite{mandel1979sub}, optimize single-photon sources based on, e.g., spontaneous parametric down-conversion~\cite{davis2022improved,sempere-llagostera2022reducing} or improve the efficiency of schemes like Gaussian boson sampling by enabling more precise photon-number measurements than conventional single-photon detectors~\cite{madsen2022quantum,deng2023gaussian}. 

While transition-edge sensors (TESs)~\cite{cabrera1998detection,lita2008counting} represent a state-of-the-art PNR technology, offering intrinsic photon-number resolution with high efficiencies, their operation requires millikelvin temperatures, complex readouts, and their intrinsically slow recovery time limits their experimental accessibility. 
Superconducting nanowire single-photon detectors (SNSPDs) have rapidly emerged as the leading technology for single-photon detection, owing to their excellent performance characteristics, including near-unity detection efficiency~\cite{reddy2020superconducting,chang2021detecting}, picosecond timing jitter~\cite{korzh2020demonstration}, fast reset times~\cite{muenzberg2018superconducting}, and exceptionally low dark count rates~\cite{chiles2022new}. Historically, SNSPDs have been employed as threshold detectors, distinguishing only between the absence and presence of one or more photons. Multiplexing of conventional threshold detectors (in space or time), can been used to gain quasi-PNR~\cite{paul1996photon,fitch2003photon,achilles2003fiber}.
In this context, recent work has achieved high efficiency, high maximum count rates, low dark counts and low crosstalk by multiplexing up to 64 SNSPDs into one active area~\cite{craiciu2023high,stasi2024enhanced,zhang2024superconducting,fleming2025high,ding2025photon}. However, multiplexing can only emulate true PNR with sufficiently many threshold detectors~\cite{sperling2012true}. 
As quantum systems grow, more detectors are needed, posing a significant scaling challenge. Therefore, true or intrinsic PNR is beneficial. 

Groundbreaking research revealed that SNSPDs possess an inherent photon-number-resolving capability~\cite{cahall2017multi}. Since then, models for photon-number resolution have been constructed~\cite{nicolich2019universal,sidorova2025jitter} and different PNR readout techniques have been investigated based on signal amplitude via impedance-matching tapers~\cite{zhu2020resolving,colangelo2023impedance,kong2025a} or signal timing characteristics of rising (and falling) edges~\cite{endo2021quantum,sempere-llagostera2022reducing,davis2022improved,sauer2023resolving,schapeler2024electrical,kong2024large,los2024high,jaha2024kinetic}.

Fundamentally, the intrinsic PNR capability of SNSPDs arises from the formation of multiple resistive regions (hotspots) on the superconducting nanowire upon absorption of multiple photons.
When multiple photons are absorbed within a sufficiently short time window, they create multiple distinct hotspots, and the total resistance of the nanowire becomes proportional to the number of absorbed photons. Signatures of PNR should still be resolvable up to the resistive domain lifetime~\cite{sidorova2025jitter}, which is roughly $50~\mathrm{ps}$ for NbTiN SNSPDs~\cite{sidorova2024low}.

In this work, we aim to advance the fundamental understanding of operating SNSPDs as PNR detectors, since they offer a promising pathway for numerous cutting-edge quantum applications that demand precise photon-number resolution with high speed and low noise.
To do so, we investigate the intrinsic PNR capabilities of a commercial SNSPD based on a detailed analysis of arrival-time histograms derived from time differences between a trigger signal and the rising edge of the SNSPD's electrical signal~\cite{sauer2023resolving,schapeler2024electrical,sidorova2025jitter}.
We specifically examine the photon-number differentiation under varying temporal optical pulse widths, which significantly influences resolution. 
We compare a recently constructed model for arrival-time histograms based on exponentially-modified Gaussian distributions~\cite{sidorova2025jitter} with a simpler model based on Gaussian distributions, focusing on misidentification of photon numbers. 
Finally, to provide a complete and rigorous characterization of the detector performance, we reconstruct the Positive-Operator-Valued Measures (POVMs) of the PNR SNSPD. 

\section{Experimental Setup}\label{sec:setup}
\begin{figure}[ht]
    \centering
    \includegraphics[width=1\linewidth]{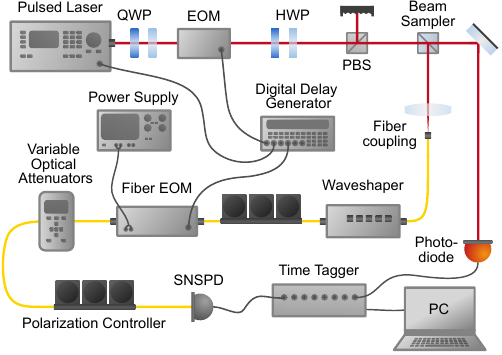}
    \caption{Experimental setup used to investigate the influence of multiple experimental parameters of the input light pulses on the photon-number resolution of SNSPDs. The laser pulses from a pulsed laser pass through two electro-optic modulators for high extinction ratio pulse-picking, a programmable optical processor (waveshaper) to manipulate the temporal shape of the pulse and variable optical attenuators to set the mean photon number per pulse. For more detail refer to the main text in Sec.~\ref{sec:setup}. QWP: Quarter-wave plate; HWP: half-wave plate; EOM: electro-optic modulator; PBS: polarizing beam splitter.}
    \label{fig:setup}
\end{figure}

The experimental setup is shown in Fig.~\ref{fig:setup}. We use a $1550~\mathrm{nm}$ pulsed laser with a repetition rate of $80~\mathrm{MHz}$. 
An electro-optic modulator (EOM) is used as a pulse picker to achieve a repetition rate of $9.5~\mathrm{kHz}$ of the laser pulses. The EOM is modulated by a square-pulse from a digital delay generator, which is triggered by an electrical synchronization pulse from the pulsed laser. 
A quarter- and half-wave plate are used to rotate the polarization of the laser to horizontal for optimal pulse picking. Another set of wave plates are used after the EOM to optimize the extinction ratio of the pulse-picking, which was maximally $23~\mathrm{dB}$. The unpicked pulses are dumped at a beam dump, after a polarizing beam splitter (PBS). The transmitted pulses are further split by a beam sampler, where the transmitted part is used to trigger a fast photodiode with a bandwidth of $5~\mathrm{GHz}$, and the reflected part is fiber-coupled.
We use a programmable optical processor (WaveShaper 4000B from Coherent), to investigate different pulse shapes and pulse widths. 
Afterwards, the light passes through a fiber polarization controller to optimize the polarization for another (fiber-coupled) EOM to enhance the extinction ratio of the pulse picking, achieving over $50~\mathrm{dB}$. 
Two variable optical attenuators (VOAs) are used to precisely set the mean photon number per pulse and another set of fiber polarization controllers are used to optimize the efficiency of the polarization-dependent SNSPD. The commercial SNSPD (Single Quantum) has a jitter of $19~\mathrm{ps}$ (FWHM) and is biased at saturated internal detection efficiency. 
A time tagger (Time Tagger X from Swabian Instruments) is used to record the time stamp of the trigger signal from the fast photodiode as well as the rising and falling edge of the electrical trace of the SNSPD. The threshold of the time tagger is set to half peak height, which was previously found to give optimal separation of photon-number events~\cite{schapeler2024electrical}.
Calibration of the input states is performed following the procedure outlined in Refs.~\cite{schapeler2020quantum,schapeler2024electrical}, which relates the mean photon number per pulse of the coherent states with the attenuation of the variable optical attenuators. 

For the main data set used for Sec.~\ref{sec:analysis}, we record a total of $D=122$ different coherent input states (with an optical temporal pulse duration of $2.9~\mathrm{ps}$). We vary the mean photon number per pulse $\bar{n}$ linearly from $0$ to $16$ in steps of $1$, then quadratically from $25$ to $10000$, then with linear attenuation steps up to a maximum mean photon number of approximately $68000$. Every input state is sampled $570000$ times; we record the timestamps of the trigger signal and rising and falling edge of the SNSPD trace. 

\section{Pulse-duration dependent PNR}\label{sec:methods}
\begin{figure}
    \centering
    \includegraphics[width=1\linewidth]{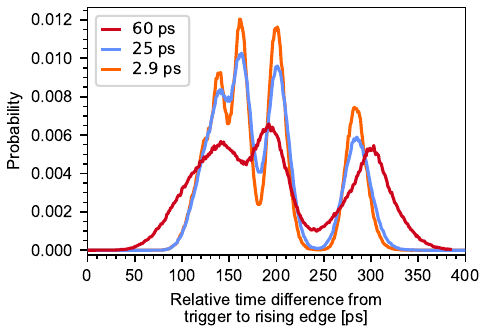}
    \caption{Arrival-time histogram of different waveshaper bandwidth settings, for a mean photon number per pulse of $\bar{n}=3$ and center wavelength of $\lambda=1550~\mathrm{nm}$. The temporal pulse widths of $60~\mathrm{ps}$ (red), $25~\mathrm{ps}$ (light blue) and $2.9~\mathrm{ps}$ (orange) correspond to bandwidths of $0.01~\mathrm{nm}$, $0.14~\mathrm{nm}$ and $2.66~\mathrm{nm}$, respectively. The mean photon number for the bandwidth of $0.01~\mathrm{nm}$ is 2.5, due to limited optical power.}
    \label{fig:filterBandwidth}
\end{figure}

The intrinsic photon-number resolving capability of an SNSPD depends on the total measured jitter in the experimental setup~\cite{sidorova2025jitter}. One contribution to this jitter is the optical temporal pulse duration of the laser pulses, i.e, the uncertainty of the photon arrival time. We investigate the PNR capability of the SNSPD for varying optical pulse durations by shaping the pulse in the spectral domain (see Appendix~\ref{sec:pulseShaping}). The effect of different spectral filters on the PNR capability is described in Appendix~\ref{sec:filterShape} and visualized in Fig.~\ref{fig:filterType}; all following measurements use a Gaussian filter, as this leads to a cleaner differentiation of photon-number events. 
In Fig.~\ref{fig:filterBandwidth} we show arrival-time histograms for different bandwidth settings, i.e., optical temporal pulse durations. The bandwidth of $2.66~\mathrm{nm}$ (orange line) leads to the calculated optical pulse duration of $2.9~\mathrm{ps}$, which gives the cleanest photon-number differentiation.  
The light blue line in Fig.~\ref{fig:filterBandwidth} shows that the photon-number differentiation starts to decrease for an optical pulse duration of $25~\mathrm{ps}$ (a filter bandwidth of $0.14~\mathrm{nm}$). 
If one of the contributions to the system jitter begins to dominate (in this case the optical temporal pulse duration), the underlying distributions for all photon-number contributions begin to broaden; this is clearly visible by the increased one-photon contribution (compare rightmost peaks in Fig.~\ref{fig:filterBandwidth}).
For an optical pulse duration of $60~\mathrm{ps}$, the separation between the one and two photon contributions is significantly affected (see peaks at relative time differences of $310~\mathrm{ps}$ and $180~\mathrm{ps}$ of the red line in Fig.~\ref{fig:filterBandwidth}, respectively).

\section{Analysis}\label{sec:analysis}
For every incident mean photon number on the detector, we analyze the data as arrival-time histograms between the trigger signal and the rising edge of the SNSPD trace. With these histograms we investigate the quality of photon-number assignment. We note that although we also recorded the falling edge of the SNSPD trace, we found that for this data set no additional photon-number information could be extracted~\cite{footnote1}, as shown in Refs.~\cite{sauer2023resolving,schapeler2024electrical}. 
Subsequently, we will describe the fitting procedure of the arrival-time histograms.

\subsection{Exponentially-modified Gaussian distributions} \label{sec:fittingEMGs}
Arrival-time histograms from SNSPDs are well described by exponentially-modified Gaussian (EMG) distributions~\cite{sidorova2017physical}. Recently, Ref.~\cite{sidorova2025jitter} extended this into a model incorporating photon-number resolution based on arrival-time measurements, including the underlying jitter mechanism in superconducting nanowires. 

\begin{figure*}[ht]
    \centering
    \includegraphics[width=1\linewidth]{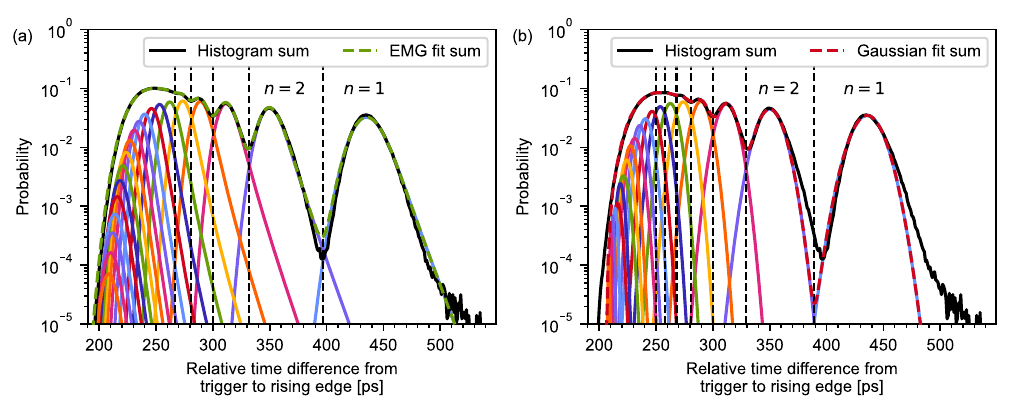}
    \caption{(a) Sum of the arrival-time histograms (black line) with fits (green dashed line and the colored lines) based on EMG distributions (for an ensemble of arrival-time histograms for different mean photon numbers), according to the model from Ref.~\cite{sidorova2025jitter} with $\chi^2_\mathrm{EMG}=0.02$. (b) Sum of the arrival-time histograms (black line) with a fit (red dashed line) based on a sum of Gaussian distributions weighted according to Poissonian statistics (Eq.~\ref{eqn:gaussianSum}) with $\chi^2_\mathrm{Gauss}=0.06$. 
    Vertical black dashed lines separate the photon-number regions, based on the intersection points between neighboring distributions. Individual underlying Gaussian distributions for photon-number contributions up to $n=20$ are shown with different colored lines.}
    \label{fig:gaussfit}
\end{figure*}

Our analysis based on EMG distributions uses the first nine input states $\bar{n}=1$ to $\bar{n}=9$ to incorporate a slight shift in the peak positions of the underlying distributions when varying the mean photon number (further detail can be found in Ref.~\cite{sidorova2025jitter}; the slight shift can be seen in Appendix~\ref{sec:fittingGaussians} in Fig.~\ref{fig:histsum}(a), colored lines). This ensures the analysis is independent of mean photon number. We fit EMG distributions (based on the model from Ref.~\cite{sidorova2025jitter}) to the histograms for different mean photon numbers and sum the individual fits. Using this model, we achieve high accuracy between data and fit, as can be seen in Fig.~\ref{fig:gaussfit}(a), where the exponential tail, that is present in the experimental data, is accounted for.
By incorporating the slight shift in the one-photon peak position, we can fully characterize the quality of the photon-number assignment in a later step. This becomes important in experiments where photon numbers need to be distinguished on a shot-by-shot basis, but the mean photon number may vary.

Depending on the SNSPD material or design, the exponential contribution in the arrival-time measurements may differ. Alternatively, fitting models based on sums of Gaussian distributions have been used~\cite{los2024high,jaha2024kinetic}. We also apply a Gaussian model (which is explained in detail in Appendix~\ref{sec:fittingGaussians}), to show the difference in accuracy between both models.
Figure~\ref{fig:gaussfit}(b) shows that the fit (red dashed line) has reduced overlap with the data (black line). Plotting the underlying Gaussian distributions on a logarithmic scale also reveals further details about how the overlap decreased (colored lines).
The EMG model (Fig.~\ref{fig:gaussfit}(a)) is accurate over four orders of magnitude, whereas the Gaussian model (Fig.~\ref{fig:gaussfit}(b)) is accurate for merely one order of magnitude.

\subsection{Assignment quality} \label{sec:assignmentQuality}
In order to quantify the assignment quality of the photon-number events, we extract fitting parameters of the underlying distributions (for both the EMG and Gaussian distribution model). 
We then calculate the intersection points of all neighboring distributions, which is where we draw separation lines to distinguish the photon-number components into regions of arrival times (see vertical dashed lines in Fig.~\ref{fig:gaussfit}).

\begin{figure*}
    \centering
    \includegraphics[width=1\linewidth]{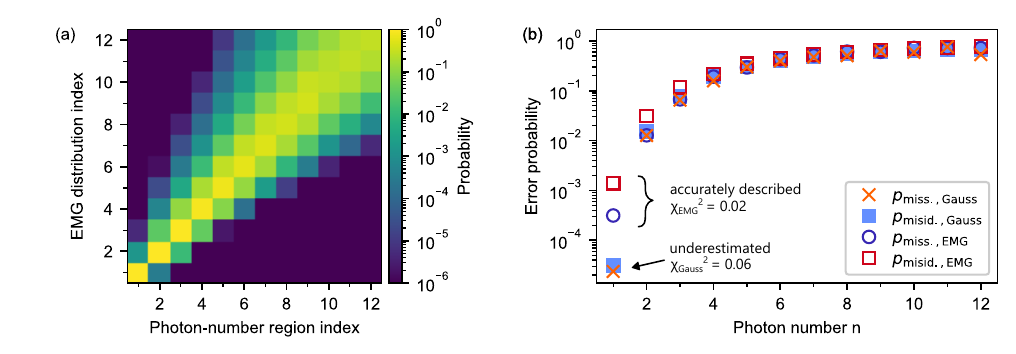}
    \caption{(a) Overlap matrix of every EMG distribution with every photon-number region, bounded by the separation lines, i.e., intersection points of the underlying distributions (from Fig.~\ref{fig:gaussfit}(a)). (b) Probability of missing $p_{\mathrm{missing,}n}$ and misidentifying $p_{\mathrm{misidentified,}n}$ an $n$-photon event as a function of photon number $n$ for both fitting methods (based on Gaussian and EMG distributions). For the Gaussian model, the probabilities are underestimated due to inaccuracies in the fit of the exponential tail of the underlying distributions (see Fig.~\ref{fig:gaussfit}(b)).}
    \label{fig:overlap}
\end{figure*}

For each fitting method, we calculate the overlap of every distribution with every photon-number region (bounded by the upper and lower bounds), which results in an overlap-matrix. The overlap matrix based on the EMG distributions is shown in Fig.~\ref{fig:overlap}(a), where the color axis is displayed on a logarithmic scale.

Subsequently, we use the type I and type II errors~\cite{dekking2005basic}, namely the probabilities to misidentify (false positive) or miss (false negative)~\cite{tiedau2020single} an $n$-photon event to quantify the assignment of photon-numbers. The probability of missing an $n$-photon event, which happens if the $n$th distribution is not fully contained within the $n$th photon-number region is given by
\begin{equation}\label{eqn:p_missing}
    p_{\mathrm{missing,}n} = 1-\int_{l_n}^{u_n}g_n\,,
\end{equation}
where $g_n$ is the $n$th distribution and $u_n$ and $l_n$ are the upper and lower bounds of the $n$th photon-number region.
The probability of misidentifying other photon-numbers $m\neq n$ as $n$, which happens if other distributions $m\neq n$ overlap with the $n$th photon-number region is given by
\begin{equation}\label{eqn:p_misidentified}
    p_{\mathrm{misidentified,}n} = \frac{\sum_{m\neq n}\int_{l_n}^{u_n}g_m}{\sum_{i}\int_{l_n}^{u_n}g_i}\,,
\end{equation}
which is normalized to all contributions in the $n$th photon-number region. The probabilities for both fitting methods are shown in Fig.~\ref{fig:overlap}(b), where it can be seen that due to the larger overlap of the underlying distributions the probabilities of missing and misidentifying an $n$-photon event increase as a function of $n$.  

The error probabilities are generally larger for the fits based on the EMG distributions (i.e., the model from Ref.~\cite{sidorova2025jitter}), due to capturing the exponential tail in the experimental data.
The simplified fits based on Gaussian distributions lead to an underestimation of the error probabilities, which can be seen most clearly in Fig.~\ref{fig:overlap}(b) for a photon number of $n=1$ (exact probabilities are listed in Table~\ref{tab:errors}).
For unity efficiency of the detector, based on the Gaussian fits one would assume to misidentify only one out of roughly 30000 one-photon events (which would actually be a two-photon event), whereas for the more accurate EMG fits one out of roughly 700 one-photon events is misidentified. This shows that the exponential tail contributes significantly to misidentification, which further motivates making the separation between photon-number components as large as possible to reduce the misidentification probability.

Alternatively, it is possible to reduce the misidentification probability $p_{\mathrm{misidentified},n}$ by choosing narrower photon-number regions around the respective peak centers. This technique has also been used to reduce assignment errors for photon counting using TESs~\cite{eaton2023resolution}. The drawback of reducing the size of the region is that ``loss'' is introduced. When the photon-number regions do not cover the entire arrival-time axis (see Fig.~\ref{fig:acceptanceWindow}(a)) some detection events lead to invalid assignments, if they do not fall into any region. In Appendix~\ref{sec:miniMisi} we show that sacrificing $6\%$ of one-photon events to loss can reduce the misidentification probability of one-photon events to $0.01\%$, i.e., one out of 10000 one-photon events is misidentified.

Finally, we use the resolvability criterion defined in Ref.~\cite{schapeler2025optimizing}, to determine the maximal photon number the SNSPD can resolve, i.e., the photon number $n$ for which the inequality
\begin{equation}\label{eqn:res_fwhm}
    (\mu_n-\mu_{n+1}) \geq \mathrm{FWHM}_n
\end{equation}
is still valid. Here $\mu_n$ is the peak center of the $n$th peak in the arrival time histogram and $\mathrm{FWHM}_n$ is the full-width-half-maximum of the $n$th photon-number contribution. According to this criterion the SNSPD can resolve up to $n=3$ photons (for both analysis methods), every event contributing to earlier times in the arrival-time histogram may be regarded as a ``four or more photon'' event. 

\begin{table}[]
\caption{Probabilities to miss or misidentify photon-number components based on the two analysis methods using either Gaussian distributions or EMG distributions.}
\label{tab:errors}
\begin{tabular}{l|l|l|l|l}
    & \multicolumn{2}{c|}{$p_\mathrm{misidentified}$} & \multicolumn{2}{c}{$p_\mathrm{missing}$} \\ \cline{2-5}
    & Gaussian & EMG & Gaussian & EMG \\ \hline
$n=1$ & $0.0032\%$ & $0.14\%$ & $0.0024\%$ & $0.032\%$ \\ \hline
$n=2$ & $1.53\%$ & $3.14\%$ &  $1.27\%$ & $1.28\%$    
\end{tabular}
\end{table}

\begin{figure*}[ht]
    \centering
    \includegraphics[width=1\linewidth]{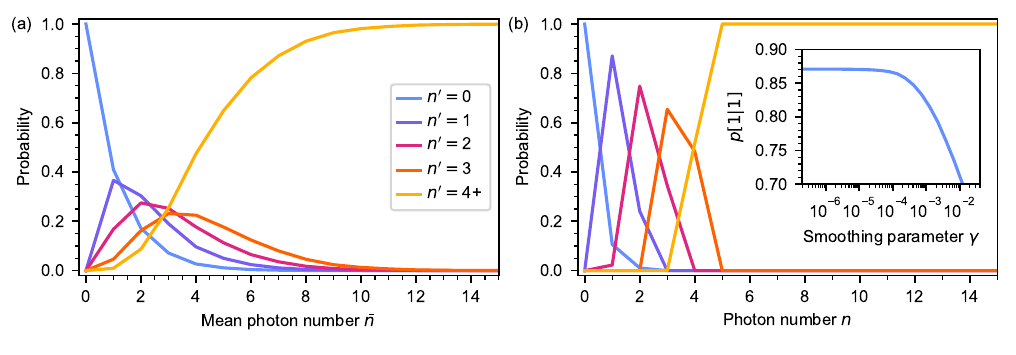}
    \caption{(a) Outcome matrix $\mat{P}$ that results from applying the separation lines from Fig.~\ref{fig:gaussfit}(a) to the recorded data, by creating histograms with variable bin sizes for the possible detector outcomes. The largest outcome is $n'=3$, given by the resolvability criterion Eq.~\ref{eqn:res_fwhm}. (b) Reconstructed POVM elements for a smoothing parameter of $10^{-6}$. The inset shows the $\Pi_{1,1}=p(1|1)$ POVM element as a function of the smoothing parameter. Over-smoothing is manifested in a reduction of the $p(1|1)$ probability for a larger smoothing parameter.}
    \label{fig:PandPOVM}
\end{figure*}

\section{Quantum detector tomography}
Outcomes of the photon-number resolving SNSPD are allocated based on their arrival-times, with the separation lines defining different outcomes (see vertical dashed lines in Fig.~\ref{fig:gaussfit}(a)). This is done by counting the occurrences in each photon-number region (i.e., creating a histogram with variable bin sizes) for every input state. The $n'=0$ outcome is populated by subtracting the number of all $n'>0$ outcomes from the total number of trials (i.e., 570000). Subsequently, the entire matrix is normalized, i.e., divided by the total number of trials, resulting in the outcome matrix $\mat{P}$ with elements $P_{d,n'}$ (see Fig.~\ref{fig:PandPOVM}(a)). Here, $d$ indexes the known coherent input states, which are also represented in a matrix $\mat{F}$ with elements $F_{d,n}=\frac{|\alpha_d|^{2n}}{n!}\mathrm{e}^{-|\alpha_d|^{2}}$ for photon numbers $n\in[0,M-1]$ and $|\alpha|^{2}=\bar{n}$ as the mean photon number of a coherent state. $M$ is the Hilbert space dimension, which describes the maximum photon-number space where the detector was operated. We truncate the matrices $\mat{P}$ and $\mat{F}$ after $d=40$ input states, as the detector is already saturated (only responds with the largest outcome, i.e. a ``four or more photon'' event); this can be seen by the yellow solid line in Fig.~\ref{fig:PandPOVM}(a) reaching a probability of one.

Finally, to reconstruct the set of positive operator valued measures (POVMs) $\{\mat{\pi}_{n'}\}$, commonly represented in a matrix $\mat{\Pi}$, that describes our (phase-insensitive) detector in a quantum mechanically consistent way, we make use of the Born rule in matrix representation, namely  
\begin{equation}\label{eqn:POVM}
    \mat{P}=\mat{F}\mat{\Pi}\,.
\end{equation}
For that we solve the constrained minimization problem~\cite{lundeen2009tomography, feito2009measuring,schapeler2020quantum,schapeler2025scalable}
\begin{equation}\label{eqn:minimization}
    \mathrm{min}\bigl\{||\mat{P}-\mat{F}\mat{\Pi}||_2+\gamma\sum_{n,n'}\left(\Pi_{n,n'}-\Pi_{n+1,n'}\right)^2\bigr\}\,,
\end{equation}
which is subject to the constraints $\mat{\pi}_n\succeq0$ and $\sum_{n'=0}^{N-1}\mat{\pi}_{n'}=\mat{I}$ and where $\gamma$ is a smoothing parameter. The diagonal elements of the reconstructed POVMs for the different outcomes of the detector (corresponding to the columns of the matrix $\mat{\Pi}$) are shown in Fig.~\ref{fig:PandPOVM}(b) for a smoothing parameter of $\gamma=10^{-6}$. To find the optimal smoothing parameter, we sweep $\gamma$ and investigate the matrix element $\Pi_{1,1}=p(n'=1|n=1)$, i.e., the conditional probability to observe the $n'=1$ outcome given a one-photon input in the photon-number basis (see the inset in Fig.~\ref{fig:PandPOVM}(b)). We choose a value that does not affect the $\Pi_{1,1}$ element, to prevent over-smoothing. 

The POVMs (based on a resolution limit of $n=3$ from Eq.~\ref{eqn:res_fwhm}) in Fig.~\ref{fig:PandPOVM}(b) have sharp features. This is expected due to the intrinsic photon-number resolution and high efficiency of the SNSPD. For a photon-number resolving detector, the outcomes of a POVM reconstruction should directly map to photon numbers, i.e., the basis is the photon-number (Fock) basis, excluding the largest outcome of ``n or more photons''. When setting the resolvability limit of the PNR detector too high (i.e., there is too much overlap of the underlying photon-number contributions in the photon-number regions of the arrival-time histogram), the reconstructed POVMs cannot represent the photon-number basis anymore, at least not for larger outcomes. Thus, the outcomes need to be seen as labels, describing events that occur in a specific arrival-time window, rather than photon numbers.

\begin{figure*}[ht]
    \centering
    \includegraphics[width=1\linewidth]{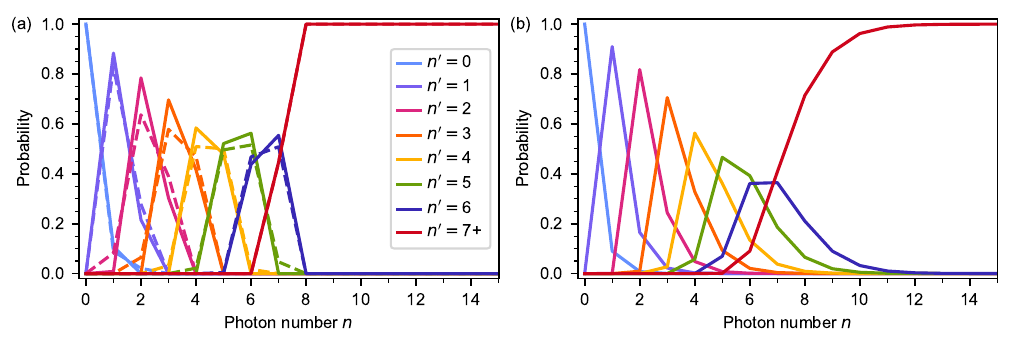}
    \caption{(a) Reconstructed POVMs $\mat{\Pi}_\mathrm{rec}$ when allowing the largest outcome to be $n'=6$. The outcomes remain sharp even for larger $n'$. Solid lines show the reconstructed POVMs for $\gamma=10^{-6}$ and dashed lines for $\gamma=10^{-3}$. (b) Simulated POVMs $\mat{\Pi}_\mathrm{sim}$ using the overlap matrix $\mat{O}$ in Fig.~\ref{fig:overlap}(a) and the loss matrix $\mat{L}$ (Eq.~\ref{eqn:lossmatrix}). The outcomes broaden as $n'$ increases.}
    \label{fig:POVMcomparison}
\end{figure*}

Using the resolvability criterion (Eq.~\ref{eqn:res_fwhm}) leads to the expected sharp features of the reconstructed POVMs (shown in Fig.~\ref{fig:PandPOVM}(b) and Table~\ref{tab:POVMs}), as we limit the photon-number differentiation based on arrival times to reasonably low overlaps between neighboring photon-number contributions.
In order to investigate the POVMs of the detector for a case where the overlap between the underlying photon-number contributions is large, we allow the detector to resolve up to $n=6$ photons, i.e., the highest outcome is a ``seven or more photon'' event. Following the procedure of the previous POVM reconstruction, we find the POVMs shown in Fig.~\ref{fig:POVMcomparison}(a).
Based on the individual EMG distributions in Fig.~\ref{fig:gaussfit}(a) and the overlap matrix in Fig.~\ref{fig:overlap}(a), the larger outcomes $n'>3$ are sharper than expected. As more photon numbers contribute to a photon-number region with increasing $n$, the POVMs should broaden (become smoother) for larger outcomes.

In an effort to visualize the expected behavior of the POVMs, we model the POVMs of the detector as $\mat{\Pi}=\mat{L}\mat{O}$, using the overlap matrix $\mat{O}$ from Fig.~\ref{fig:overlap}(a) and the loss matrix $\mat{L}$. The elements of $\mat{L}$ are given by~\cite{kruse2017limits}
\begin{equation}\label{eqn:lossmatrix}
    L_{n,m} = \binom{n}{m}\eta^m(1-\eta)^{n-m}\,,
\end{equation}
which describe the probabilities to retain $m$ out of $n$ photons with an efficiency $\eta=\eta_\mathrm{det}=0.91\pm0.03$. The overlap matrix $\mat{O}$ acts similar to the convolution matrix~\cite{achilles2004photon,coldenstrodt2009proposed}, which is used for multiplexed detectors. The simulated POVMs shown in Fig.~\ref{fig:POVMcomparison}(b) behave as expected, as larger outcomes become broader due to the increasing overlap from multiple EMG distributions with the photon-number regions.

To test the accuracy of the reconstructed POVMs $\mat{\Pi}_\mathrm{rec}$ and the simulated POVMs $\mat{\Pi}_\mathrm{sim}$, we can use Eq.~\ref{eqn:POVM} to calculate back the outcome matrices $\mat{P}_\mathrm{rec}$ and $\mat{P}_\mathrm{sim}$ and compare it to the measured outcome matrix $\mat{P}$ (based on our experimental data). For the comparison we use a fidelity measure~\cite{feito2009measuring,schapeler2025scalable}
\begin{equation}\label{eqn:fidelity}
    F = \frac{\left(\sum_i\sqrt{a_ia'_i}\right)^2}{\sum_ia_i\sum_ia'_i}\,,
\end{equation}
to quantify how well two vectors (columns) of the outcome matrices match, where $a$ and $a'$ correspond to the same column in the experimentally measured outcome matrix $\mat{P}$ and the outcome matrix using the POVMs $\mat{P}_\mathrm{rec}$ or based on the simulation $\mat{P}_\mathrm{sim}$, respectively.
Even though the POVMs in Fig.~\ref{fig:POVMcomparison} are different, we find that both $\mat{P}$ matrices, based on the reconstructed and simulated POVMs, describe the experimental data well with average fidelities of 99.92\% and 99.80\% across the non-saturated outcomes, respectively. 

In order to smoothen larger outcomes, one might consider to increase the smoothing parameter $\gamma$. However, that does not work, as all outcomes are affected by the smoothing, thus also decreasing the sharpness of outcomes $n'\leq3$. This can be seen by the dashed lines in Fig.~\ref{fig:POVMcomparison}(a), which shows the result of a reconstruction using $\gamma=10^{-3}$, with an average fidelity of 99.89\%. We find that the reconstruction with $\gamma=10^{-6}$ yields the highest average fidelity, thus giving the most accurate representation of the experimental data.

\begin{table}[]
\caption{Reconstructed POVM elements, i.e., conditional probabilities $\Pi_{n,n'}=p(n'|n)$ with a smoothing parameter of $\gamma=10^{-6}$ (from Fig.~\ref{fig:PandPOVM}(b)) up to a photon number of $n=7$.}
\label{tab:POVMs}
\begin{tabular}{cc|ccccc}
                                         &              & \multicolumn{5}{c}{Detector outcome $n'$} \\
                                         & $\Pi_{n,n'}$ & 0 & 1 & 2 & 3 & 4+ \\ \hline
\parbox[t]{2mm}{\multirow{8}{*}{\rotatebox[origin=c]{90}{Photon number $n$}}} & 0 & 0.99996 & 0.00004 & 0 & 0 & 0 \\
                                         & 1 & 0.10640 & 0.87092 & 0.02268 & 0 & 0 \\
                                         & 2 & 0.01081 & 0.24109 & 0.74810 & 0 & 0 \\
                                         & 3 & 0 & 0 & 0.34561 & 0.65439 & 0 \\
                                         & 4 & 0 & 0 & 0 & 0.48357 & 0.51643 \\
                                         & 5 & 0 & 0 & 0 & 0 & 1.00000 \\
                                         & 6 & 0 & 0 & 0 & 0 & 1.00000 \\
                                         & 7 & 0 & 0 & 0 & 0 & 1.00000 \\
\end{tabular}
\end{table}

\section{Conclusion}\label{sec:conclusion}
The assignment of photon numbers using SNSPDs depends on many experimental parameters. An important parameter is the jitter of the system, which influences the resolution of arrival-time measurements. Minimizing the jitter components of an experimental setup is crucial for optimized PNR~\cite{los2024high,sidorova2025jitter}. One contribution to jitter, i.e., the uncertainty in arrival time of the photons, is the optical pulse duration of the light source. Using a programmable optical pulse shaper, we investigate different pulse shapes and pulse durations, by setting different filters and bandwidths, respectively. We find that a Gaussian spectral filter give a much cleaner arrival-time histogram compared to a bandpass filter (see Fig.~\ref{fig:filterType}). We also show that the photon-number discrimination remains accurate for temporal pulse durations in the same order of magnitude as other jitter contributions of the experiment. For $60~\mathrm{ps}$ optical pulses the discrimination is significantly reduced (see Fig.~\ref{fig:filterBandwidth}). 

Additionally, we show a comparison between two different models for photon-number measurements using SNSPDs, based on exponentially-modified Gaussian distributions and Gaussian distributions. 
At first glance an analysis of the arrival-time histograms using Gaussian distributions leads to accurate fits of the photon-number contributions (see Fig.~\ref{fig:gaussfit}(b)). However, our analysis shows that the probability to misidentify photon-number events is significantly underestimated due to the exponential tail in the arrival-time histograms (see Table~\ref{tab:errors}). Using a more accurate model for arrival-time histograms (based on a sum of EMG distributions~\cite{sidorova2025jitter} for the SNSPD in this investigation) leads to excellent agreement between experimental data and fit, and thus to an accurate characterization of the misidentification probability of photon-number events (see Table~\ref{tab:errors}).
Minimized misidentification probability is achieved when the photon-number contributions to the arrival-time histograms have large separation in time, or alternatively narrowing acceptance regions in the arrival times for photon-number events (see Appendix~\ref{sec:miniMisi}); this will enable high accuracy in heralding single-photons based on, e.g., spontaneous parametric down-conversion sources.

We follow our analysis with a reconstruction of the POVMs of the detector. The POVMs show sharp features, which is expected from photon-number resolving detectors, which ideally show a one-to-one mapping from input photon numbers $n$ to detector outcomes $n'$. 
We allow the detector to resolve higher photon numbers (up to $n=6$ instead of $n=3$), at the cost of increased overlap of photon-number contributions in the arrival-time histograms. We find that the POVMs remain moderately sharp, even for larger outcomes. This does not fit to the expected broadening based on simulated POVMs, due to contributions from multiple photon numbers due to higher overlap of the underlying distributions (see Fig.~\ref{fig:gaussfit}(a) and Fig.~\ref{fig:overlap}(a)). In this operation mode, the outcomes correspond to mixtures of photon numbers, which describe events that occur in a specific arrival-time window. Notwithstanding, the POVMs can still represent the data with high fidelity.

\section*{Acknowledgments}
We thank Laura Serino and Lorenzo M. Procopio for fruitful discussions about optical pulse shaping and optical pulse duration calculation; and Maximilian Protte for feedback on the manuscript. 
Partially funded by the European Union (ERC, QuESADILLA, 101042399). Views and opinions expressed are however those of the author(s) only and do not necessarily reflect those of the European Union or the European Research Council Executive Agency. Neither the European Union nor the granting authority can be held responsible for them. 
This work has received funding from the German Ministry of Education and Research within the PhoQuant project (grant number 13N16103). 
F.S. is part of the Max Planck School of Photonics supported by the German Federal Ministry of Research, Technology and Space (BMFTR), the Max Planck Society, and the Fraunhofer Society.

\section*{Data Availability Statement}
The data that support the findings of this study are openly available in Zenodo at \url{https://doi.org/10.5281/zenodo.14888684}.

\appendix
\section{Optical pulse shaping} \label{sec:pulseShaping}
The programmable optical processor (waveshaper) allows one to filter the original laser spectrum with different filter shapes and bandwidths. 
Spectral shaping of the optical pulse may also change the temporal shape of the optical pulse and for Fourier-transform-limited pulses, the temporal pulse duration can be calculated as
\begin{equation}\label{eqn:tbp}
    \Delta\tau = \mathrm{TBP}\frac{\lambda^2}{c\Delta\lambda}\,.
\end{equation}
Here $\Delta\lambda$ is the filtered spectral bandwidth (set by the waveshaper) and $\mathrm{TBP}$ is the time-bandwidth product, which is $0.441$ for Gaussian pulses and $0.315$ for sech$^2$ pulses. 
We use a free-space femtosecond-pulsed laser (with a sech$^2$ pulse shape), which is coupled into fiber. Due to dispersion in the fiber, the optical pulse is not Fourier-transform-limited when arriving at the waveshaper. We use an autocorrelator to measure the optical pulse duration of $1.06~\mathrm{ps}$ before the spectral shaping (after $5~\mathrm{m}$ of fiber in total; $1~\mathrm{m}$ fiber for fiber-coupling and $4~\mathrm{m}$ of fiber inside the waveshaper before the shaping stage).

Given limited optical power we are not able to measure the pulse duration using an autocorrelator after spectral shaping. Therefore, we calculate the optical pulse duration based on the spectral parameters. Due to group-velocity dispersion, the output pulse length $\Delta\tau_\mathrm{out}$ is given by~\cite{kruse2017limits}
\begin{equation}\label{eqn:dispersion}
\Delta\tau_\mathrm{out}=\Delta\tau_\mathrm{in}\sqrt{1+\left(\frac{2\mathrm{ln}(2)}{\pi}\frac{\lambda^2}{c'\Delta\tau_\mathrm{in}^2} D_\lambda L_\mathrm{fiber}\right)^2}\,,
\end{equation}
where $\Delta\tau_\mathrm{in}$ is the FWHM pulse length of the input pulse, $\lambda$ is the wavelength, $c'=\frac{c}{n}$ the speed of light in fiber (with $n=1.4682$~\cite{corning}), $D_\lambda$ is the dispersion coefficient and $L_\mathrm{fiber}$ is the length of fiber. The dispersion coefficient at $1550~\mathrm{nm}$ for standard SMF28 fiber is given by $D_\lambda\leq18~\frac{\mathrm{ps}}{\mathrm{nm}\,\mathrm{km}}=1.8\times10^{-5}~\frac{\mathrm{s}}{\mathrm{m}^2}$~\cite{corning}.

We use Eq.~\ref{eqn:dispersion} to calculate the optical pulse length of the photons arriving at the cryostat, accounting for approximately $L_\mathrm{fiber}=32~\mathrm{m}$ of fiber-based group-velocity dispersion. Different waveshaper bandwidths lead to different optical temporal pulse durations of our probe pulses (see Table~\ref{tab:pulselengths}). Based on Eq.~\ref{eqn:tbp} we would calculate an optical temporal pulse duration of $>300~\mathrm{ps}$ for the bandwidth of $\Delta\lambda=0.01~\mathrm{nm}$. Given the experimental result, i.e., the arrival-time histogram from Fig.~\ref{fig:filterBandwidth} (red line) and the bandwidth setting accuracy of the waveshaper of $\pm0.04~\mathrm{nm}$, an optical temporal pulse duration of $(60\pm10)~\mathrm{ps}$ is more realistic. Potentially, there are also effects from the $\mathrm{TBP}$ not being purely described by the Gaussian filter, rather by a convolution from the original sech$^2$ shape of the laser and the Gaussian filter. Bandwidths and corresponding optical temporal pulse durations are listed in Table~\ref{tab:pulselengths}.
\begin{table}[h]
\caption{Calculated optical pulse durations according to different waveshaper bandwidth settings. The uncertainties are based on the bandwidth setting accuracy of the waveshaper ($\pm0.04~\mathrm{nm}$) and the $\mathrm{TBP}$ that might be a convolution from the original sech$^2$ shape (0.315) of the laser and the Gaussian filter (0.441).}
\label{tab:pulselengths}
\begin{tabular}{l|l|l|l}
Bandwidth [nm]      & 0.01  & 0.14  & 2.66   \\ \hline
Pulse duration [ps] & $60\pm10$ & $25\pm10$ & $2.9\pm0.3$
\end{tabular}
\end{table}

\section{Dependence on filter shape} \label{sec:filterShape}
\begin{figure}[t]
    \centering
    \includegraphics[width=1\linewidth]{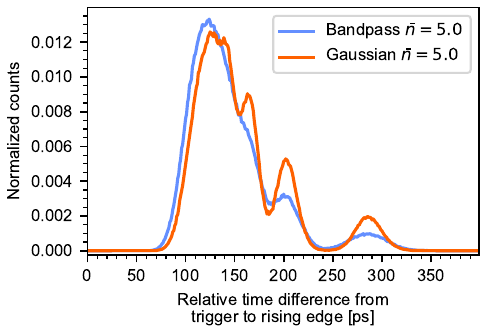}
    \caption{Arrival-time histogram given two waveshaper filter types, for a mean photon number per pulse of $\bar{n}=5$, bandwidth of $\Delta\lambda=0.14~\mathrm{nm}$ and center wavelength of $\lambda=1550~\mathrm{nm}$.}
    \label{fig:filterType}
\end{figure}
Figure~\ref{fig:filterType} shows a comparison of the arrival-time histogram when setting a bandpass filter (light blue) and a Gaussian filter (orange) with a bandwidth of $\Delta\lambda=0.14~\mathrm{nm}$ for the waveshaper. 
The Gaussian filter leads to a much cleaner photon-number differentiation, as the bandpass filter seems to wash out the photon-number events. 
Applying a bandpass or Gaussian filter in the spectral domain leads to a sinc-like or Gaussian temporal pulse shape, respectively. With this in mind, it explains that the Gaussian filter leads to a better performance, as the side lopes of the sinc temporal shape likely cause the pulse to be broader in time.

\section{Gaussian distributions} \label{sec:fittingGaussians}
\begin{figure*}[ht]
    \centering
    \includegraphics[width=1\linewidth]{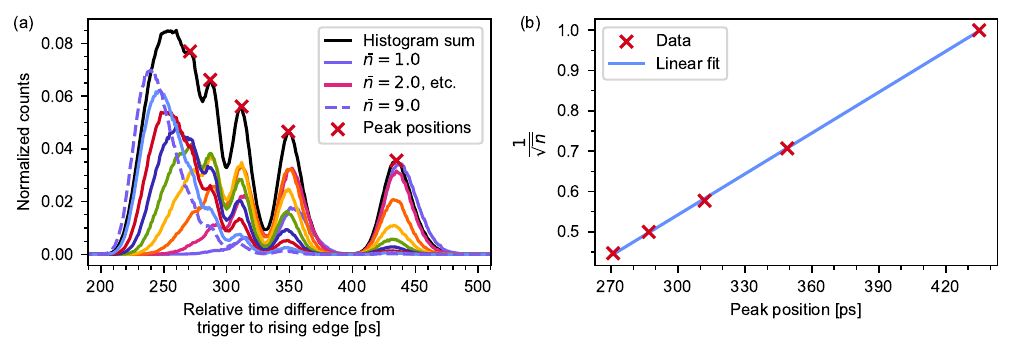}
    \caption{(a) Arrival-time histograms for mean photon numbers from $\bar{n}=1$ to $\bar{n}=9$ (colored lines are scaled up by a factor of three, to improve visibility) as well as the sum of the histograms (black line). Peak positions of the first five photon-number contributions are marked by red crosses. (b) Predicted $1/\sqrt{n}$ scaling~\cite{nicolich2019universal} of the peak positions from (a) with a linear fit (blue line).}
    \label{fig:histsum}
\end{figure*}

In Fig.~\ref{fig:histsum}(a) it can be seen that the arrival-time histograms for different incident mean photon numbers, specifically the one-photon peak position, slightly shifts to earlier times with increasing mean photon number. 
To incorporate this slight shift, we sum the histograms for incident mean photon numbers from one to nine (in steps of one) and find the peak position of the first five photon-number components.

According to Nicolich et al.~\cite{nicolich2019universal} the rise time of the trace scales with $1/\sqrt{n}$; this matches our data, which can be seen by the linear dependence when plotting $1/\sqrt{n}$ over the peak position of the $n$th photon-number contribution (see Fig.~\ref{fig:histsum}(b)). 
We then fit the histogram sum with a sum of Gaussian distributions
\begin{equation}\label{eqn:gaussianSum}
    f(a,\sigma_n,x_n) = a\cdot\sum_{n=1}^{20}\frac{1}{\sigma_n\sqrt{2\pi}}\mathrm{e}^{-\frac{1}{2}\left(\frac{x-x_n}{\sigma}\right)^2}P'(n)\,,
\end{equation}
with $a$ as an overall scaling factor, $\sigma_n$ as the standard deviations and $x_n=\left(\frac{1}{\sqrt{n}}-b_\mathrm{lin}\right)/m_\mathrm{lin}$ as the peak position of the Gaussian distributions. Here we use $b_\mathrm{lin}$ and $m_\mathrm{lin}$ from the linear fit in Fig.~\ref{fig:histsum}(b) as fitting parameters to enforce the $1/\sqrt{n}$ scaling. Each Gaussian distribution is scaled according to the sum of the Poisson distribution components from the mean photon numbers used in the histogram sum, i.e., $P'(n)=\sum_{\bar{n}=1}^9\frac{(\bar{n}\eta)^n}{n!}\mathrm{e}^{-\bar{n}\eta}$, where $\eta=0.91$ is the efficiency of the detector.

\section{Minimizing misidentification} \label{sec:miniMisi}
\begin{figure*}[ht]
    \centering
    \includegraphics[width=1\linewidth]{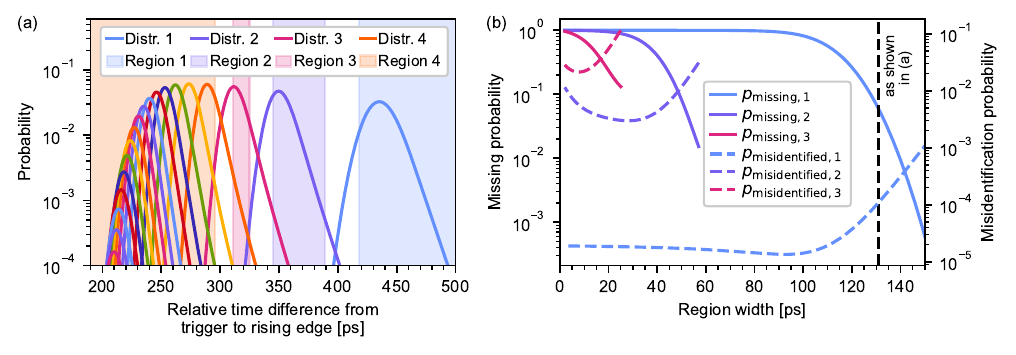}
    \caption{(a) Arrival-time histogram fits based on EMG distributions (equivalent to Fig.~\ref{fig:gaussfit}(a)) with photon-number regions optimized to simultaneously reduce $p_{\mathrm{misidentified},n}$ and $p_{\mathrm{missing},n}$. The overlap between (EMG) distributions and regions (with the same color) is maximized, while the overlap of any region with all other (EMG) distributions is minimized. (b) Probabilities of missing (left axis, solid lines) and misidentification (right axis, dashed lines) as a function of region width for $n=1$, $n=2$ and $n=3$. As an example, we leave the right border of the region fixed as shown in (a) and shrink the region from the left border, starting from the intersection point of the distributions. The actual optimization process is more complex, as the region can also shrink from the right edge simultaneously.}
    \label{fig:acceptanceWindow}
\end{figure*}

It is possible to come up with a minimization scheme that minimizes the misidentification probability $p_{\mathrm{misidentified},n}$, while simultaneously minimizing the probability to miss events $p_{\mathrm{missing},n}$; in other words, the overlap of the $n$th distribution with the $n$th region should be maximized, while the overlap of all other distributions with the $n$th region should be minimized simultaneously. Choosing the weights of the two tasks sensibly (i.e., deciding which case has higher priority for an experiment) will optimize the bounds for the photon-number regions. The probabilities to misidentify or miss events are shown in Table~\ref{tab:acceptanceWindow} for the optimized photon-number regions shown in Fig.~\ref{fig:acceptanceWindow}(a). We visualize the probabilities of missing and misidentification as a function of region width in Fig.~\ref{fig:acceptanceWindow}(b). To create Fig.~\ref{fig:acceptanceWindow}(b), we leave the right border of each photon-number region fixed (as shown in Fig.~\ref{fig:acceptanceWindow}(a)) and shrink the region from the left border, starting from the intersection point of the distributions. Sacrificing roughly $p_{\mathrm{missing},1}\approx6\%$ of one-photon events, will reduce the misidentification probability to $p_{\mathrm{misidentified},1}=0.01\%$ (see Fig.~\ref{fig:acceptanceWindow}(b) at a region width of $131~\mathrm{ps}$), i.e., only one out of roughly 10000 one-photon events are actually a two-photon event.

\begin{table}[th]
\caption{Probabilities to miss or misidentify photon-number components based on EMG distributions and optimized photon-number regions from Fig.~\ref{fig:acceptanceWindow}(a).}
\label{tab:acceptanceWindow}
\begin{tabular}{l|l|l}
      & $p_\mathrm{misidentified}$ & $p_\mathrm{missing}$ \\ \hline
$n=1$ & $0.01\%$                   & $5.87\%$         \\
$n=2$ & $0.60\%$                   & $24.30\%$        \\
$n=3$ & $2.78\%$                   & $48.85\%$        \\             
\end{tabular}
\end{table}

\newpage

\bibliography{references}

\end{document}